\documentclass[a4paper,11pt]{article}

\usepackage{jcappub} 

\usepackage[T1]{fontenc}

\title{Constraining the dynamical dark energy parameters: Planck-2013 vs WMAP9}

\author[a]{B. Novosyadlyj,}
\author[a,d]{O. Sergijenko,}
\author[b]{R. Durrer,}
\author[c]{V. Pelykh}

\affiliation[a]{Astronomical Observatory of Ivan Franko National University of Lviv,\\ Kyryla i Methodia str., 8, Lviv, 79005, Ukraine}
\affiliation[b]{Universit\'e de Gen\`eve, D\'epartement de Physique Th\'eorique and CAP,\\ 24 quai Ernest-Ansermet, CH-1211 Gen\`eve 4, Switzerland}
\affiliation[c]{Ya. S. Pidstryhach Institute for Applied Problems of Mechanics and Mathematics,\\ Naukova str., 3-b, Lviv, 79060, Ukraine}
\affiliation[d]{Main Astronomical Observatory of the National Academy of Sciences of Ukraine, \\ Zabolotnoho str., 27, Kyiv, 03680, Ukraine}

\emailAdd{novos@astro.franko.lviv.ua}
\emailAdd{olka@astro.franko.lviv.ua}
\emailAdd{ruth.durrer@unige.ch}
\emailAdd{pelykh@iapmm.lviv.ua}

\abstract{
We determine the best-fit values and confidence limits for dynamical dark energy parameters together with other cosmological parameters on the basis of different datasets which include WMAP9 or Planck-2013 results on CMB anisotropy, BAO distance ratios from recent galaxy surveys, magnitude-redshift relations for distant SNe Ia from SNLS3 and Union2.1 samples and the HST determination of the Hubble constant. We use a Markov Chain Monte Carlo routine to map out the likelihood in the multi-dimensional parameter space. We show that the most precise determination of cosmological parameters with the narrowest confidence limits is obtained for the Planck{+}HST{+}BAO{+}SNLS3 dataset. The best-fit values and 2$\sigma$ confidence limits for cosmological parameters in this case are $\Omega_{de}=0.718\pm0.022$, $w_0=-1.15^{+0.14}_{-0.16}$, $c_a^2=-1.15^{+0.02}_{-0.46}$, $\Omega_bh^2=0.0220\pm0.0005$, $\Omega_{cdm}h^2=0.121\pm0.004$, $h=0.713\pm0.027$,  $n_s=0.958^{+0.014}_{-0.010}$, $A_s=(2.215^{+0.093}_{-0.101})\
cdot10^{-9}$, $\tau_{rei}=0.093^{+0.022}_{-0.028}$.  For this dataset, the 
$\Lambda$CDM model is just outside the 
2$\sigma$ confidence region, while for the dataset WMAP9{+}HST{+}BAO{+}SNLS3 the $\Lambda$CDM model is only 1$\sigma$ away from the best fit. The tension in the determination of some cosmological parameters on the basis of two CMB datasets WMAP9 and Planck-2013 is highlighted.
}

\begin{document}
\maketitle
\flushbottom

\section{Introduction}

The year 2013 is epochal for cosmology owing to the publication of observational data from two space observatories, the final WMAP results (at the end of 2012) \cite{wmap9a,wmap9b} and first cosmological Planck results \cite{Planck2013a,Planck2013b,Planck2013c}. WMAP has started the precision epoch of cosmology and Planck has a good chance to improve it considerably. Cosmologists are trying to use these data to answer important questions especially about the dark sector of the Universe. In this paper we study the nature of dark energy. 

Most observational data obtained so far can not distinguish between a cosmological constant and
quintessence/phantom dark energy at a statistically significant level (see e. g. books
\cite{Amendola2010,Wolschin2010,Ruiz2010,Novosyadlyj2013m}, special issue of the journal General Relativity and Gravitation \cite{GRG2008} and citations therein): the marginalized $1\sigma$ range of the equation of state (EoS) parameter $w_{de}$ as a rule covers a comparable interval on both sides of $w_{\Lambda}=-1$. 
We have recently analyzed \cite{Novosyadlyj2013} the arbitrating power of available and expected observational data in distinction between quintessence and phantom types of dark energy and have studied  prospects to improve this situation. The data on CMB temperature anisotropy from the Planck satellite and the parameters of baryon acoustic oscillations (BAO) extracted from advanced galaxy surveys seem to be most promising.

One of the models of dark energy, suitable for the problem of recognizing of its type, is a scalar field with
$w_{de}<-1/3$ that fills the Universe almost uniformly. One of the simplest scalar field models of dynamical dark energy that can be quintessence, vacuum or phantom is the scalar field with barotropic EoS \cite{Novosyadlyj2010,Sergijenko2011,Novosyadlyj2012}. The existence of analytical solutions for the evolution of such a scalar field, their regularity and applicability for any epoch in the past as well as in the future make it a useful model to investigate the general properties of dark energy, especially to establish its type -- quintessence, vacuum or phantom.  

The goal of this paper is the estimation of parameters of scalar field dark energy jointly with other cosmological
parameters on the base of different datasets including either WMAP9 or Planck CMB anisotropy measurements, as well as the last data releases on BAO and SNe Ia magnitude-redshift relations.   The paper is a sequel of Ref.~\cite{Novosyadlyj2013}, where a similar analysis has been preformed with pre-Planck data, and we have found that the data cannot measure the adiabatic sound speed $c_a^2$ and therefore cannot distinguish between quintessence and phantom models.  Here we find that with Planck data $c_a^2$ can be measured and phantom models are somewhat preferred (at the $2\sigma$ level).

\section{Scalar field model of dark energy and cosmological background}

We assume the standard paradigm of inflationary cosmology: the Universe is filled with baryonic (b) matter, cold dark matter (cdm), dark energy (de), neutrinos ($\nu$) and cosmic microwave background (CMB) radiation (r); its structure is formed by gravitational instability  from nearly scale invariant primordial perturbations generated during inflation. The post-inflationary evolution of the Universe including the dynamics of expansion, cosmological nucleosynthesis and recombination, formation of CMB anisotropies and large scale structure is well elaborated \cite{Ma1995,Durrer2001,Novosyadlyj2007,Durrer2008,Amendola2010,Wolschin2010,Ruiz2010,Novosyadlyj2013m} and implemented numerically \cite{cmbfast96,cmbfast99,Doran2005,camb,camb_source,Lesgourgues2011,Blass2011,class_source} for computation of theoretical predictions with subpercent accuracy. The model of dark energy has to be specified. 
   
We assume a scalar field model of minimally coupled dark energy which describes equally well quintessence and phantom dark energy with constant or variable EoS parameters. There exist different methods for specifying a scalar field which can mimic these properties (see books cited above). In this paper we consider the scalar field model with generalized linear barotropic EoS $p_{de}=c_a^2\rho_{de}+C$ \cite{Babichev2005,Holman2004}, where $c_a^2\equiv\dot{p}_{de}/\dot{\rho}_{de}$ and $C$ are arbitrary constants which define the dynamical properties of the scalar field on a cosmological background, which is assumed to be spatially flat, homogeneous and isotropic with Friedmann-Robertson-Walker (FRW) metric $ds^2=c^2dt^2-a^2(t)\delta_{\alpha\beta} dx^{\alpha}dx^{\beta}$, where $a(t)$ is the scale factor normalized to 1 today and we set $c=1$. The differential equation of energy-momentum conservation law $T^i_{0;i}=0$ for dark energy with such EoS leads to Riccati ordinary differential equation for $w_{de}$
\begin{equation}
\frac{dw_{de}}{d\ln{a}}=3(1+w_{de})(w_{de}-c_a^2), \label{dwda}
\end{equation}
which for $c_a^2=const$ has the analytical solution  
\begin{equation}
w_{de}=\frac{(1+c_a^2)(1+w_0)}{1+w_0-(w_0-c_a^2)a^{3(1+c_a^2)}}-1, \label{w_de} 
\end{equation}
where $w_0$ is the EoS parameter at the current epoch, $a=1$. This allows to obtain an analytical solution for the density $\rho_{de}$ 
\begin{equation}
\rho_{de}=\rho_{de}^{(0)}\frac{(1+w_0)a^{-3(1+c_a^2)}+c_a^2-w_0}{1+c_a^2}, \label{rho_de}
\end{equation} 
where $\rho_{de}^{(0)}$ is the density of dark energy at the current epoch. Clearly, $\rho_{de}$ and $p_{de}$ are analytical functions of $a$ for any values of the constants $c_a^2$ and $w_0$, which define\footnote{The constant $C$ is connected with them by the relation $C=\rho_{de}^{(0)}(w_0-c_a^2)$.} the type and the dynamics of the scalar field. They are the parameters of our dark energy model which must be determined jointly with other cosmological parameter. Both have a clear physical meaning: $w_0$ is the EoS parameter $w_{de}$ at the current epoch, $c_a^2$ is asymptotic value of the EoS parameter $w_{de}$ at early times ($a\rightarrow 0$) for
$c_a^2>-1$ and in the far future ($a\rightarrow \infty$) for $c_a^2<-1$. The  asymptotic value of $w_{de}$ in the opposite time direction is $-1$ in both cases. 

\begin{figure}
\includegraphics[width=.5\textwidth]{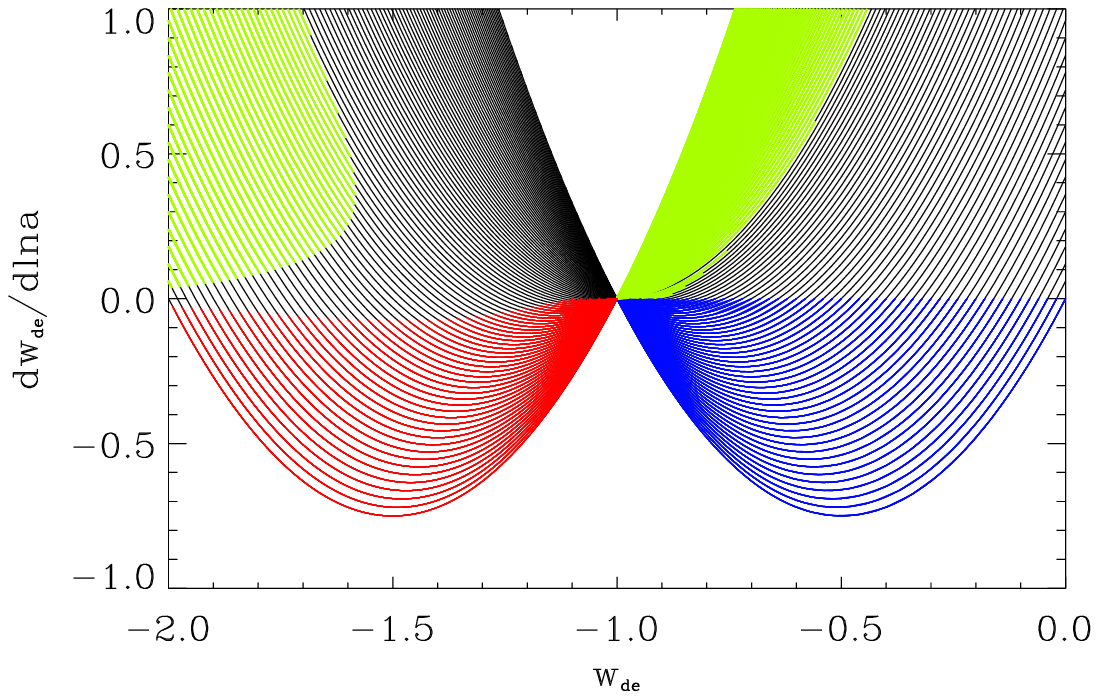}
\includegraphics[width=.5\textwidth]{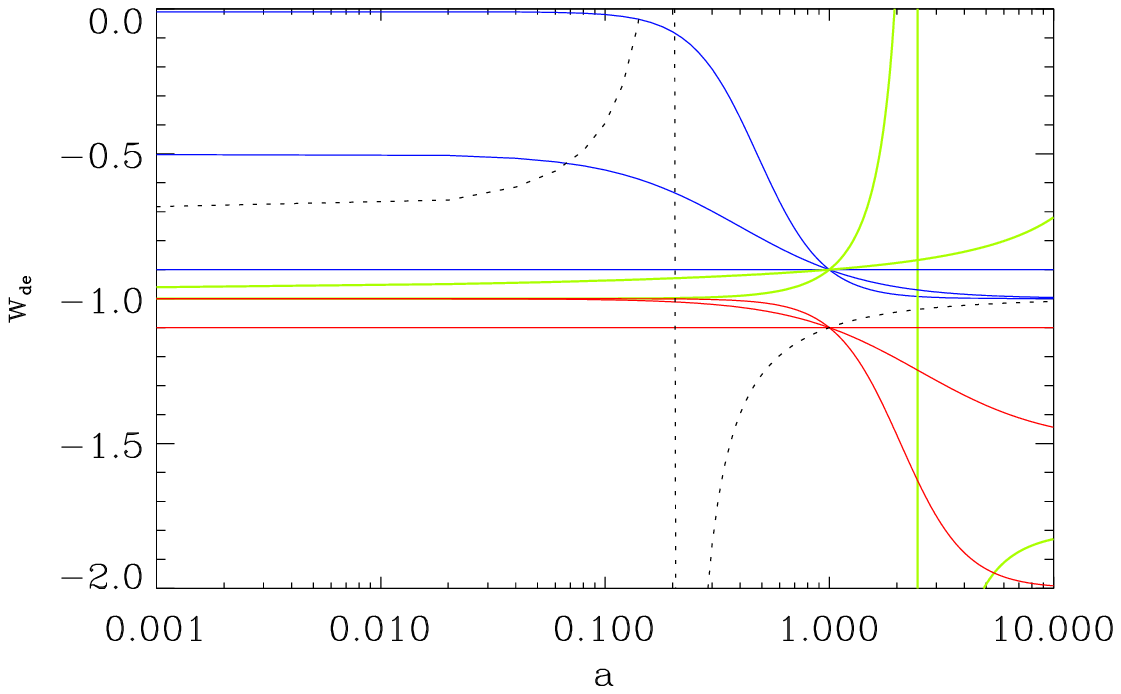}
\caption{Left: phase plane $dw_{de}/d\ln{a}-w_{de}$  (\ref{dwda}) for dark energy models with $-2\le c_a^2\le0$. Colors denote the regions occupied by models with $-1.1\le w_{de}\le -0.9$ at the current epoch: blue -- models with decreasing $w_{de}$ and raising repulsion (freezing quintessence dark energy), green -- models with increasing $w_{de}$ and receding repulsion (thawing quintessence dark energy, which becomes ``false phantom'' one with $w_{de}<-1$ but $\dot{\rho}_{de}<0$, $\rho_{de}<0$, $p_{de}>0$ in far future for short time in the vicinity of turnaround point, see \cite{Novosyadlyj2010} for details), red -- phantom models with decreasing $w_{de}$ and raising repulsion of dark energy. Right: examples of evolution tracks of EoS parameter $w_{de}$ from the colored ranges on the left. The dotted line represents the subclass of dark energy models with $w_0<-1$ and $c_a^2>w_0$, for which $\rho_{de}<0$ at some time in the past and which is excluded from further analysis.}
\label{dw-w}
\end{figure}

The phase plane of Eq.~(\ref{dwda}) and examples of evolution tracks of EoS parameter $w_{de}$ for quintessence (freezing and thawing \cite{Caldwell2005}) and phantom dark energy are shown in figure \ref{dw-w}. For more details of dynamical features of such scalar field models of dark energy see \cite{Caldwell2005,Novosyadlyj2010,Novosyadlyj2012}.

The Friedmann equations for our model of the Universe are
\begin{eqnarray}
 H&\equiv&\frac{\dot{a}}{a}=H_0\sqrt{\Omega_r a^{-4}+\Omega_m a^{-3}+\Omega_{de}f(a)}, \label{H} \\
q&\equiv&-\frac{a\ddot{a}}{\dot{a}^2}=\frac{1}{2}\frac{2\Omega_r a^{-4}+\Omega_m a^{-3}+(1+3w_{de})\Omega_{de}f(a)}
{\Omega_r a^{-4}+\Omega_m a^{-3}+\Omega_{de}f(a)},\label{q}
\end{eqnarray}
where $\dot{ }\equiv d/dt$, $f(a)=\rho_{de}/\rho_{de}^{(0)}$, $H_0$ is Hubble constant and $\Omega_r\equiv \rho_{r}^{(0)}/\rho_{tot}^{(0)}$, $\Omega_m\equiv \rho_{m}^{(0)}/\rho_{tot}^{(0)}$, 
$\Omega_{de}\equiv \rho_{de}^{(0)}/\rho_{tot}^{(0)}$ are the dimensionless density parameters for radiation, matter and dark energy components correspondingly at the current epoch ($\rho_{tot}^{(0)}\equiv \rho_{r}^{(0)}+\rho_{m}^{(0)}+\rho_{de}^{(0)}$).   The matter density parameter is the sum of cold dark matter, baryons and active neutrinos, $\Omega_m\equiv\Omega_{cdm}+\Omega_b+\Omega_{\nu}$. We follow~\cite{Planck2013c}, which includes a minimal-mass normal hierarchy for the neutrino masses: a single massive eigenstate with $m_{\nu}=0.06$ eV. It gives very small contributions into current matter density component, $\Omega_{\nu}\approx m_{\nu}/93.04h^2$ eV $\approx0.0006/h^2$, where $h\equiv H_0/100$km/s$\cdot$Mpc. The current density of thermal radiation (CMB)  is also very small, $\Omega_r=16\pi Ga_{SB}T_0^4/3H_0^2=2.49\cdot10^{-5}h^{-2}$, where $T_0$ is the current CMB temperature assumed here and below to be $2.7255$K. The first Friedmann equation (\ref{H}) today yields the constraint $\Omega_r +\
Omega_m+ \Omega_{de}=1$ (vanishing curvature). 

The scalar field  causes accelerated expansion when $(1+3w_{de})\Omega_{de}f(a) +\Omega_m a^{-3}<0$ and defines the future of the Universe. Here $f(a)=\rho_{de}(a)/\rho_{de}(1)$. 

\begin{figure}
\includegraphics[width=.5\textwidth]{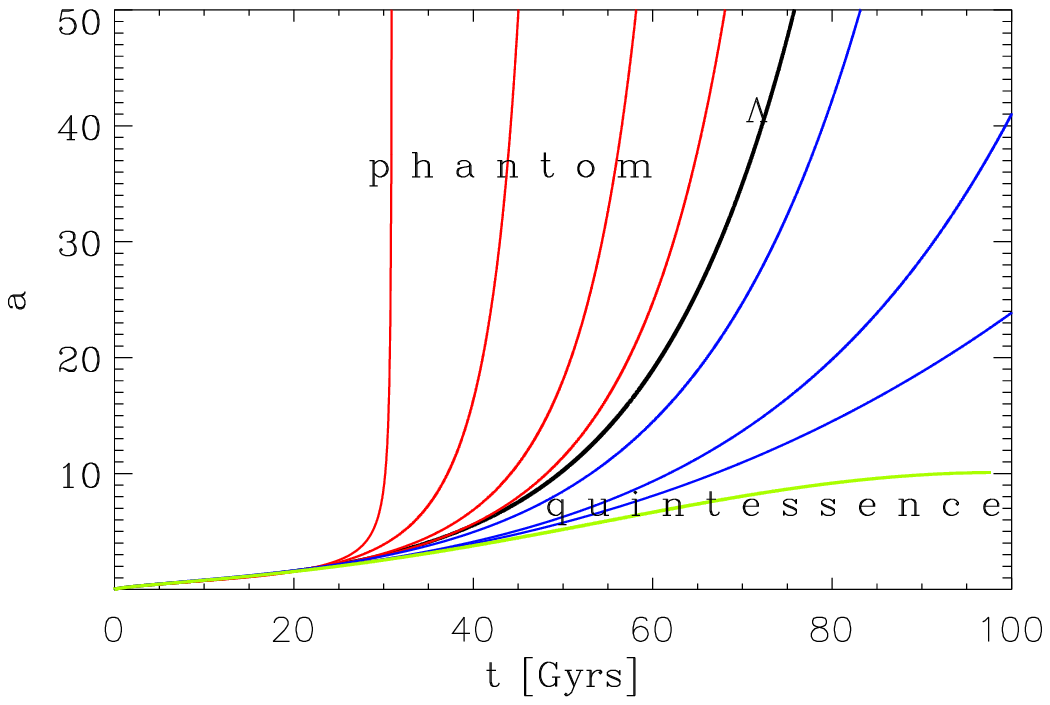}
\includegraphics[width=.5\textwidth]{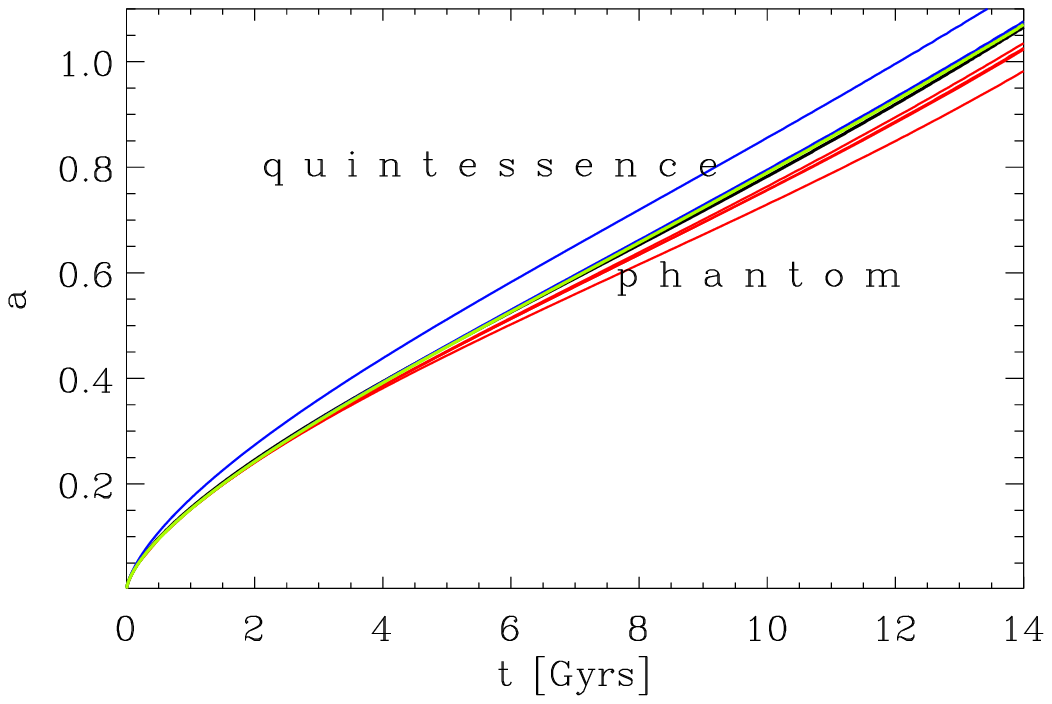}
\caption{Possible dependence of the scale factor on time, $a(t)$, for cosmological models with barotropic  scalar field dark energy. The colors correspond to models of the corresponding color in Fig.~\ref{dw-w}. The right panel corresponds to the lower left corner of the left panel. In the left panel at $t>25$ Gyrs, the models really start to differ from each other. The smaller $w_{de}$, the faster the phantom models head towards a 'Big Rip'. The green line represents 'thawing quintessence' and heads towards a 'Big Crunch' singularity. (The dependences are calculated for models with $H=70$ km/s/Mpc, $\Omega_m=0.3$, $\Omega_{de}=0.7$ and different sets of values of $w_0$ and $c_a^2$: 1 -- (-1.2, -2.0), 2 -- (-1.4, -1.2), 3 -- (-1.1, -1.2), 4 -- (-1.03, -1.2), 5 -- (-1.0, -1.0), 6 -- (-0.8, -0.0), 7 -- (-0.8, -0.7), 8 -- (-0.8, -0.8) and 9 -- (-0.8, -0.9) from top left to bottom right in the left panel. In the right panel they are in the inverse order relatively to the line of $\Lambda$-model (5, black line): 
6, 7,
 8, 9, 5, 4, 3, 2, 1.)}
\label{a-t}
\end{figure}

Integrating (\ref{H}) over $a$ we can obtain the $a-t$ relation for the model,
\begin{equation}
t(a)=\int_0^a\frac{da'}{a'H(a')}.
\label{t-a}
\end{equation}
The results for nine sets of values $w_0$ and $c_a^2$ with all other cosmological parameters fixed at the values ($H_0=70$ km/s$\cdot$Mpc, $\Omega_m=0.3$, $\Omega_{de}=0.7$) are shown in fig. \ref{a-t} (colors of lines corresponds to colors of models in fig. \ref{dw-w}). 
One sees that a scalar field with (\ref{w_de})-(\ref{rho_de}) can yield, least qualitatively, most possible scenarios of post-inflationary dynamics of expansion, predicted by modern cosmology. They are distinguished mainly by future of the Universe: eternal exponential expansion (black line in fig. \ref{a-t}), eternal power law expansion (blue lines), Big Crunch (BC) singularity (green line) and Big Rip (BR) singularity (red lines). The time of BC singularity can be estimated as twice the turn around time, $t_{BC}=2t_{ta}$, and $t_{ta}$ is the integral (\ref{t-a}) with upper limit $a_{ta}\approx [(1+w_0)/(w_0-c_a^2)]^{1/3(1+c_a^2)}$ \cite{Novosyadlyj2010,Sergijenko2011}. For example, the model indicated by the green line reaches a BC singularity in 186 Gyrs ($a_{ta}\approx10$). The BR singularity is reached for all models with a phantom scalar field within finite time estimated as \cite{Novosyadlyj2012}
\begin{equation}
 t_{BR}-t_0\approx \frac{2}{3}\frac{1}{H_0}\frac{1}{|1+c_a^2|}\sqrt{\frac{1+c_a^2}{(1+w_0)\Omega_{de}}}\label{t_br}.
\end{equation}
For the phantom models in Fig. \ref{a-t} it will be reached approximately in 25, 35, 80 and 500 Gyrs respectively (for the red lines from left to right). 

To analyze the gravitational instability of the scalar field in the context of the formation of structure in the Universe, we need to know the effective sound speed $c_s^2=\delta p_{de}/\delta\rho_{de}$. For a given Lagrangian it is easily computed, since $c^2_s=\mathcal{L}_{,X}/(\mathcal{L}_{,X}+2X\mathcal{L}_{,XX})$, where $\mathcal{L}_{,X}\equiv\partial\mathcal{L}/\partial X$. Here $X=\frac{1}{2}(\nabla\phi)^2$ denotes the kinetic term of the scalar field. For example, for a canonical Lagrangian $\mathcal{L}_{de}=\pm X-U$ (``$+$'' for quintessence, and  ``$-$'' for phantom) the effective sound speed is equal to the speed of light.  

We assume that the large scale structure of the Universe is formed from Gaussian, adiabatic scalar perturbations generated in the early Universe. The initial power spectrum of density perturbations of all components is a power-law, $P_i(k)=A_sk^{n_s}$, where $A_s$ and $n_s$ are the amplitude and the spectral index ($k$ is wave number) which must be determined jointly with other cosmological parameters. Also the scalar field is perturbed, the system of linear differential equations for the evolution of quintessence and phantom scalar field perturbations and their numerical solutions are studied in 
Refs.~\cite{Putter2010,Novosyadlyj2012,Novosyadlyj2010,Sergijenko2011}. The main conclusions are as follows: (i) the amplitude of scalar field density perturbations at any epoch depends strongly on the parameters of the barotropic scalar field $\Omega_{de}$, $w_0$, $c_a^2$ and $c_s^2$; (ii) although the density perturbations of dark energy at the current epoch are significantly smaller than matter density perturbations, they leave noticeable imprints in the matter power spectrum, which can be used to constrain the scalar field parameters.

\section{Observational data and method}

Using expressions (\ref{rho_de})-(\ref{H}) we can compute the ``luminosity distance - redshift`` or ``angular diameter distance - redshift'' relations to constrain the above mentioned parameters by comparison with the corresponding data on standard candles (supernovae type Ia, $\gamma$-ray bursts or other) and standard rulers (positions of the CMB acoustic peaks, baryon acoustic oscillations, X-ray gas in clusters or others). 
The linear power spectrum of matter density perturbations can be computed by numerical integration of the linearized Einstein-Boltzmann equations \cite{Ma1995,Durrer2001,Novosyadlyj2007,Durrer2008} using the publicly available code CAMB \cite{camb,camb_source} with the corresponding modifications for the barotropic scalar field as dark energy, 
$$P_{lin}(k)=P_i(k)T_m^2(k;\Omega_b,\Omega_{cdm},\Omega_{de},w_0,c_a^2)\,. $$
Here $T_m(k;\Omega_b,\Omega_{cdm},\Omega_{de},w_0,c_a^2)$ is the transfer function of matter density perturbations
which depends also on the  parameters listed after the semicolon. We use the modified CAMB code also to compute the angular power spectra of CMB temperature anisotropies $C^{TT}_{\ell}$ for comparison with WMAP9 and Planck data to constrain the cosmological and DE parameters mentioned above. The calculation of CMB anisotropies requires also the knowledge of the reionization history of the Universe, which depends on complicated non-linear physics of star formation. This is parameterized by the value of optical depth from the current epoch to decoupling caused by Thomson scattering of the electrons in the re-ionized plasma. It is denoted by $\tau_{rei}$ and added to the list of parameters fitted to the data. The complete list which we determine contains 9 parameters 
$$\Theta_k: \quad \Omega_{de},\,w_0,\,c_a^2,\,\Omega_b,\,\Omega_{cdm},\,H_0,\,A_s,\,n_s, \tau_{rei},$$
only 8 of which are free since we also satisfy the constraint $\Omega_b+\Omega_{cdm}+\Omega_{de}=1$ for the spatially flat cosmological model considered here. To determine their best-fit values and confidence ranges we use the following data: 
\begin{enumerate}
\item CMB temperature fluctuations angular power spectra from either WMAP9 \cite{wmap9a} or Planck-2013 results \cite{Planck2013b};
\item Hubble constant measurement from Hubble Space Telescope (HST) \cite{Riess2011} ($H$=73.8$\pm$2.4 km/(s$\cdot$Mpc));
\item BAO data from the galaxy surveys SDSS DR7 \cite{Padmanabhan2012} (1 measurement),  SDSS DR9 \cite{Anderson2012} (1 measurement),  6dF \cite{6dF} (1 measurement) (hereafter we denote all 3 measurements together as BAO);  
\item Supernovae Ia luminosity distances from either SNLS3 compilation \cite{snls3} or Union2.1 \cite{union} compilations.
\end{enumerate}

Throughout the paper we use the combination of the Planck temperature power spectrum with the WMAP9 polarization likelihood \cite{wmap9a} and refer to this CMB data combination as Planck (see for details \cite{Planck2013b}). It is obvious that WMAP9 includes the polarization  data also.

The best-fit parameters correspond to the model with maximal likelihood function
\begin{equation}
\log(L(\mathbf{x};\Theta_k))\simeq -\frac{1}{2}(x_i-x_i^{th})C_{ij}(x_j-x_j^{th})\,.
\label{likelihood}
\end{equation}
The confidence ranges correspond to the marginalized posterior functions
\begin{equation}
 P(\Theta_k;\mathbf{x})=\frac{L(\mathbf{x};\Theta_k)p(\Theta_k)}{g(\mathbf{x})}.
\label{posterior}
\end{equation}
Here $x_i$ ($i=1$, 2, ..., $N$) are observational points of the dataset considered, $x_i^{th}$ are the corresponding theoretical predictions for the model with fixed parameters $\Theta_k$, $C_{ij}$ is the error covariance matrix, $p(\Theta_k)$ is the prior for the parameter $\Theta_k$ and $g(\mathbf{x})$ is the probability distribution function of the  data (evidence).

The parameter estimation from the data is performed using the publicly available Markov Chain Monte Carlo (MCMC) code~\cite{cosmomc_source}. The detailed description of usage of MCMC for mapping out the likelihood in the multi-dimensional parameter space can be found in \cite{cosmomc}. For datasets including WMAP9 we have used the Metropolis algorithm, while for datasets including Planck -- the fast/slow blocked decorrelation scheme of \cite{Lewis2013} combined with the fast dragging method of \cite{Neal2005}. Each run has 8 chains converged to $R-1<0.01$.

\section{Results and discussion} 
To estimate cosmological parameters and especially dark energy parameters we use WMAP9 and Planck CMB anisotropy in combination with different datasets: i) alone, ii) together with HST and BAO data, iii) the complete datasets including also SNe Ia data (SNLS3 or Union2.1). 

\begin{table}[htbp]
\centering
  \caption{The best-fit values (b-f), mean values and 2$\sigma$ confidence limits (c.l.) for the parameters of our
cosmological models determined by the MCMC technique using four different observational datasets: WMAP9 (alone), Planck (with WMAP polarization), WMAP9{+}HST{+}BAO and Planck{+}HST{+}BAO.}
  \medskip\footnotesize{
  \begin{tabular}{|c|c|c|c|c|c|c|c|c|}
    \hline 
    &\multicolumn{2}{c|}{}&\multicolumn{2}{c|}{}&\multicolumn{2}{c|}{}&\multicolumn{2}{c|}{}\\   
Parameters&\multicolumn{2}{c|}{WMAP9}&\multicolumn{2}{c|}{Planck}&\multicolumn{2}{c|}{WMAP9+HST+BAO}&\multicolumn{2}{c|}{Planck+HST+BAO}\\
    &\multicolumn{2}{c|}{}&\multicolumn{2}{c|}{}&\multicolumn{2}{c|}{}&\multicolumn{2}{c|}{}\\
    \cline{2-9}
     &&&&&&&&\\
     &b-f&2$\sigma$ c.l.&b-f&2$\sigma$ c.l.&b-f&2$\sigma$ c.l.&b-f&2$\sigma$ c.l.\\   
    &&&&&&&&\\
    \hline
    &&&&&&&&\\
$\Omega_{de}$&0.659&0.648$_{-0.160}^{+0.146}$&0.737&0.675$_{-  0.123}^{+  0.125}$&0.733&$0.723_{-0.031}^{+0.029}$&0.722&0.725$_{-0.030}^{+0.027}$\\&&&&&&&&\\
$w_0$&-0.813&-0.753$_{-0.506}^{+0.423}$&-1.234&-0.959$_{-  0.505}^{+  0.629}$&-1.238&$-1.137_{-0.258}^{+0.248}$&-1.226&-1.218$_{-0.188}^{+0.199}$\\&&&&&&&&\\
$c_a^2$&-0.821&-0.957$_{-0.709}^{+0.816}$&-1.389&-1.248$_{-  0.469}^{+  0.672}$&-1.515&$-1.323_{-0.311}^{+0.668}$&-1.482&-1.411$_{-0.227}^{+0.259}$\\&&&&&&&&\\
10$\Omega_{b}h^2$&0.226&0.227$_{-0.010}^{+0.011}$&0.222&0.220$_{-  0.006}^{+  0.006}$&0.225&$0.225_{-0.009}^{+0.010}$&0.221&0.220$_{-0.004}^{+0.004}$\\&&&&&&&&\\
$\Omega_{cdm}h^2$&0.114&0.113$_{-0.009}^{+0.009}$&0.120&0.120$_{-  0.005}^{+  0.005}$&0.119&$0.117_{-0.007}^{+0.007}$&0.121&0.121$_{-0.004}^{+0.004}$\\&&&&&&&&\\
$h$&0.635&0.633$_{-0.138}^{+0.152}$&0.736&0.674$_{-  0.124}^{+  0.155}$&0.730&$0.712_{-0.045}^{+0.045}$&0.720&0.723$_{-0.039}^{+0.040}$\\&&&&&&&&\\
$n_s$&0.976&0.978$_{-0.028}^{+0.031}$&0.963&0.960$_{-  0.014}^{+  0.015}$&0.969&$0.967_{-0.023}^{+0.024}$&0.960&0.958$_{-0.012}^{+0.013}$\\&&&&&&&&\\
$\log(10^{10}A_s)$&3.100&3.092$_{-0.060}^{+0.063}$&3.084&3.089$_{-  0.048}^{+  0.051}$&3.096&$3.097_{-0.056}^{+0.060}$&3.083&3.088$_{-0.046}^{+0.051}$\\&&&&&&&&\\
$\tau_{rei}$&0.091&0.090$_{-0.028}^{+0.030}$&0.086&0.089$_{-  0.025}^{+  0.027}$&0.088&$0.086_{-0.026}^{+0.027}$&0.085&0.088$_{-0.025}^{+0.026}$\\&&&&&&&&\\
    \hline
  \end{tabular}}
  \label{tab_wp}
\end{table}
\begin{figure}[tbp]
\includegraphics[width=0.5\textwidth]{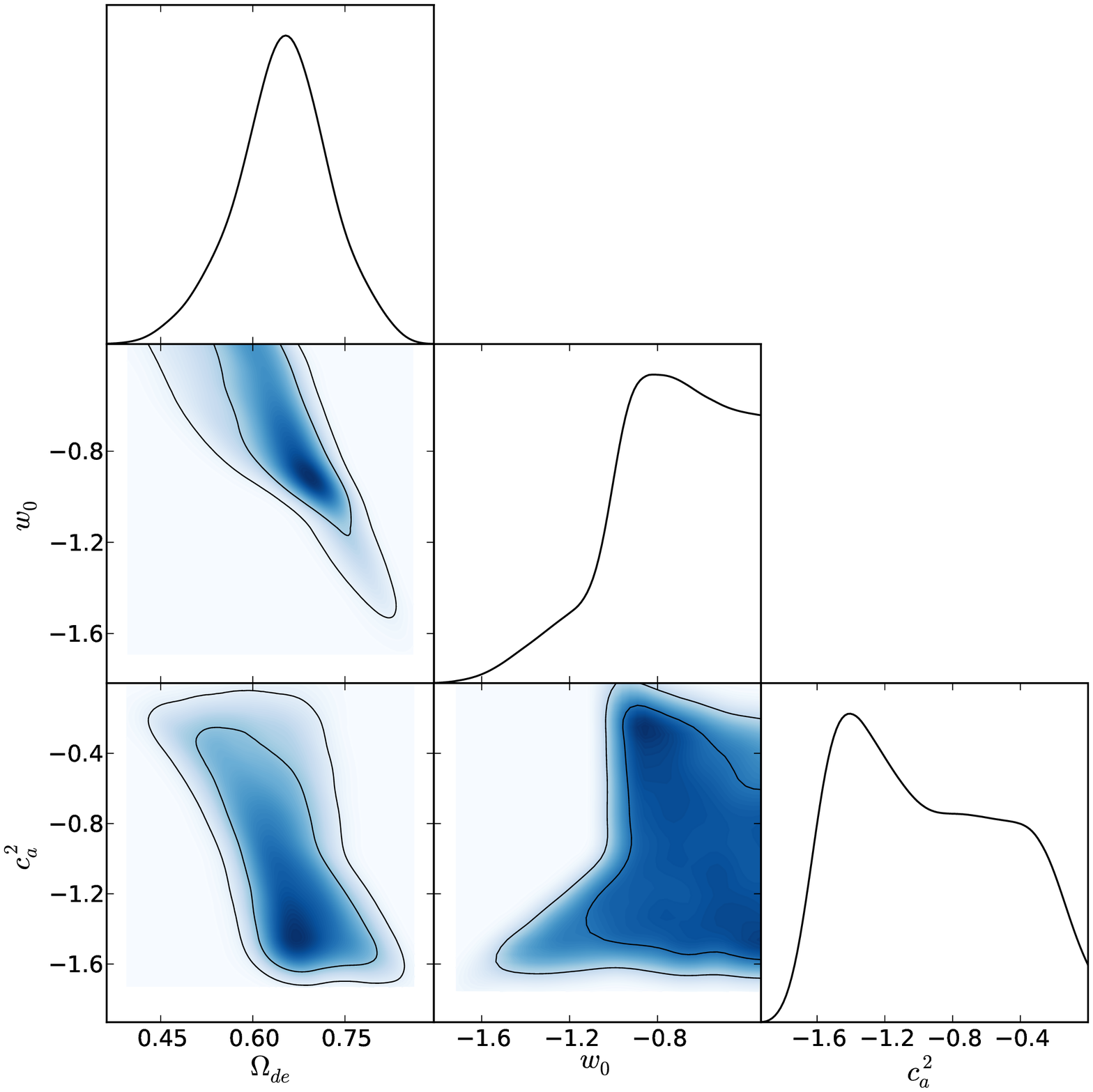}
\includegraphics[width=0.5\textwidth]{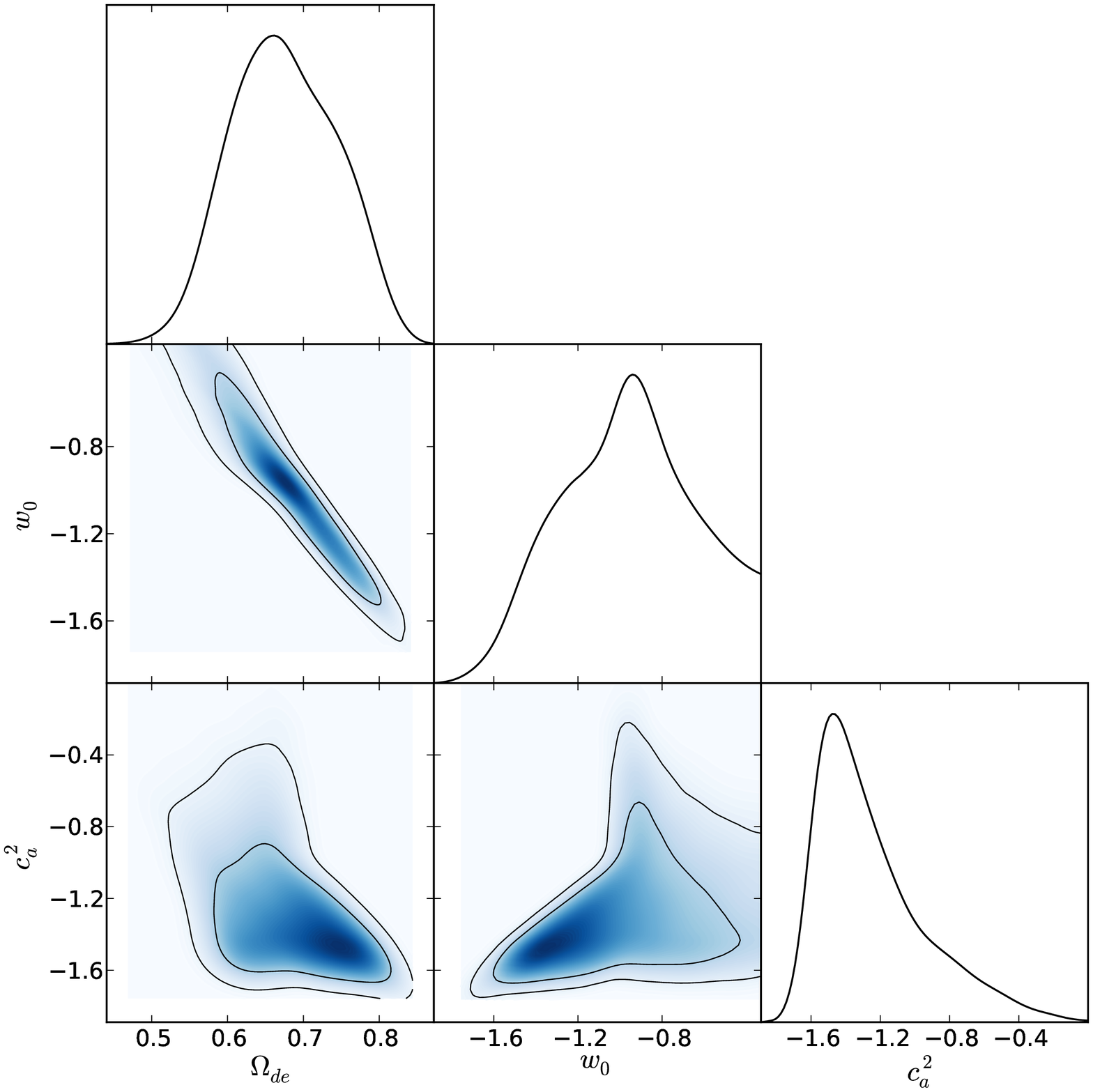}
\includegraphics[width=0.5\textwidth]{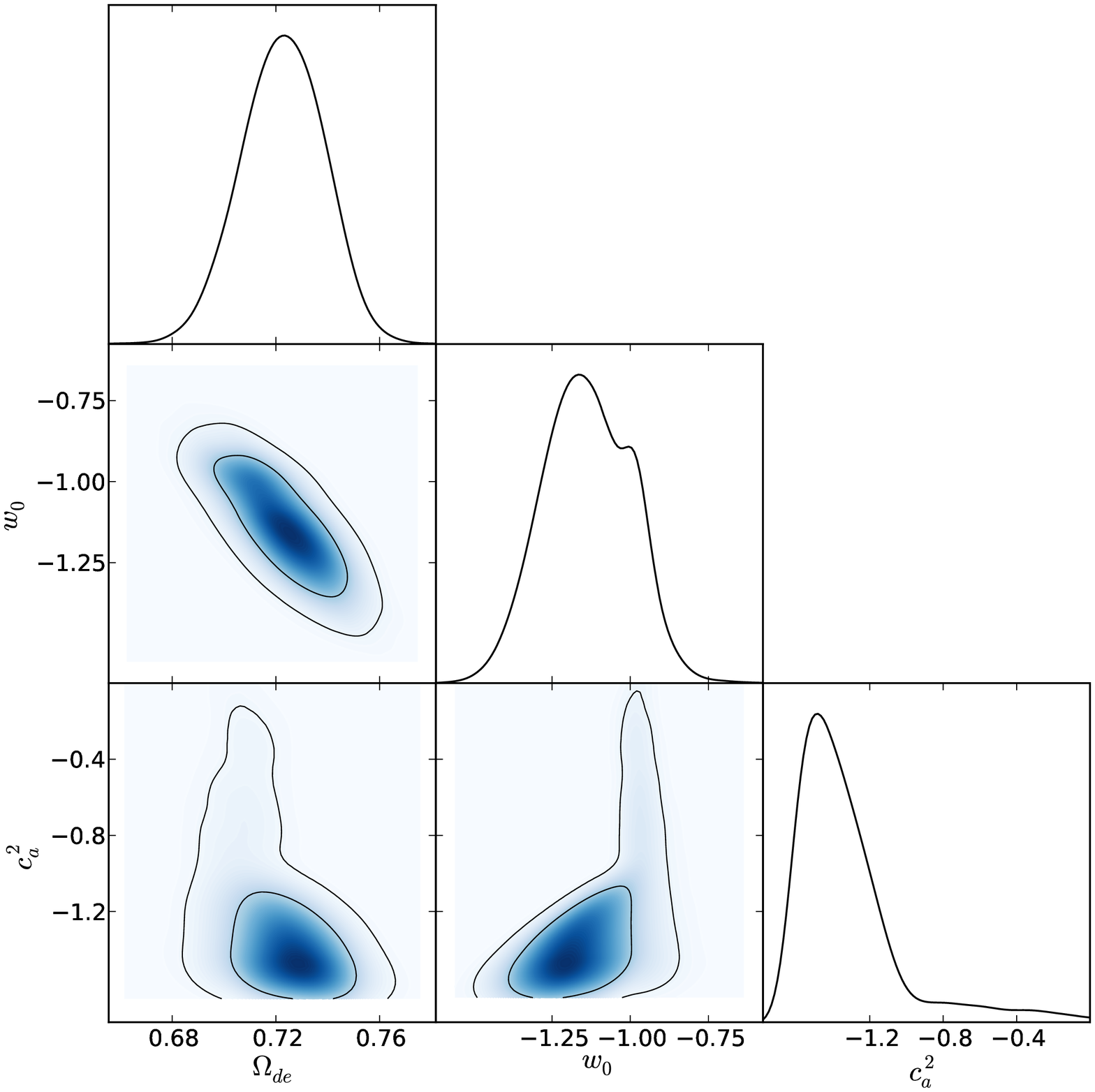}
\includegraphics[width=0.5\textwidth]{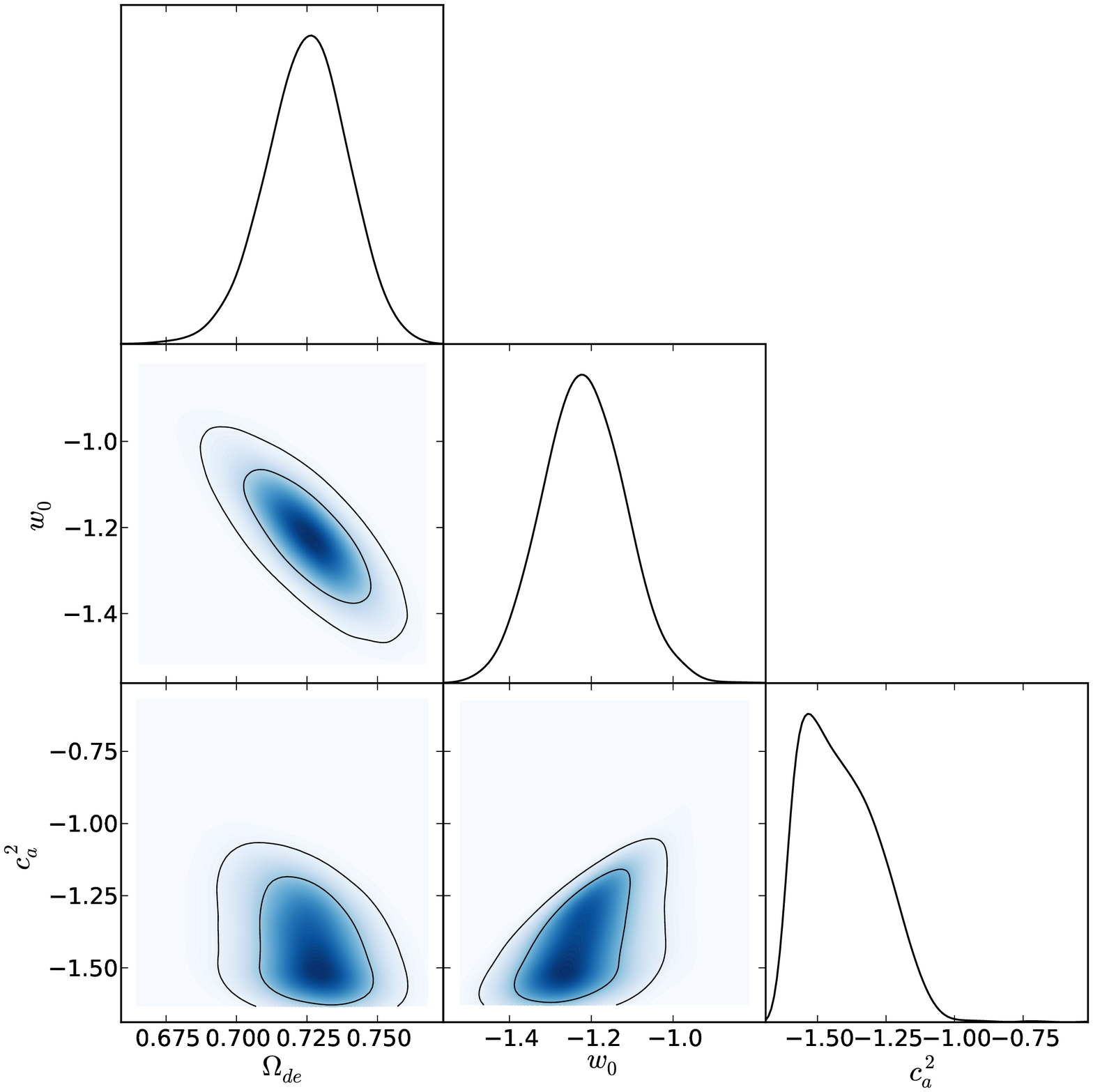}
\caption{One-dimensional marginalized posteriors (solid lines) for $\Omega_{de}$,
$w_0$ and $c_a^2$; color panels show two-dimensional marginalized posterior distributions in the planes $\Omega_{de}-w_0$, $\Omega_{de}-c_a^2$ and $w_0-c_a^2$, where solid lines show the $1\sigma$ and $2\sigma$ confidence contours. Left plots are for datasets including WMAP9 and right ones are for datasets including Planck. Top plots are for CMB alone data, bottom ones for CMB+HST+BAO data.}
\label{postlike_wp}
\end{figure}

The best-fit values of cosmological parameters, their means and the 2$\sigma$ marginalized limits  from WMAP9 or Planck CMB anisotropy data are presented in table~\ref{tab_wp} (columns 2--5) and in figure \ref{postlike_wp} (top panels). 

As expected, both CMB datasets determine the matter densities ($\Omega_bh^2$ and $\Omega_{cdm}h^2$) as well as the initial power spectrum parameters ($n_s$ and $\log(10^{10}A_s)$) with good accuracy. The accuracy of the remaining parameters is worse. The full widths of the 2$\sigma$ confidence ranges for each of these parameters in the case of WMAP9 data are 0.0021, 0.018, 0.059 and 0.123 correspondingly. For the Planck data they are narrower: 0.0012, 0.01, 0.029 and 0.099 correspondingly. The mean values are close to the best-fit ones, marginalized likelihoods and posterior functions are similar. On the other hand, 2$\sigma$ confidence ranges for $h$ and $\tau_{rei}$ determined from WMAP9 and Planck data alone are wide ($h$: 0.290 for WMAP9 and 0.279 for Planck, $\tau_{rei}$: 0.058 and 0.052 correspondingly). The dark energy parameters are rather badly constrained by CMB data alone, the marginalized likelihood (\ref{likelihood}) and posteriors (\ref{posterior}) differ significantly,
they  have different shapes and extrema. For example, the maximum of the marginalized likelihood function for $w_0$ in the case of Planck CMB data lies in the phantom range ($\approx-1.23$), while the 
maximum of the marginalized posterior function lies in the quintessence range ($\approx-0.96$). In the case of WMAP9 data both lie in the quintessence range, $\approx-0.81$ and $\approx-0.75$ correspondingly. 2D contours also illustrate the improved capabilities of Planck-2013 results to constrain the dynamical dark energy parameters as compared to  WMAP9. Our results further highlight some tension between the WMAP9 and Planck-2013 results, which is shown by fact that best-fit and mean values of the baryon and dark matter density parameters as well as of the spectral index determined from the WMAP9 data are outside the 2$\sigma$ ranges of these values determined from the Planck data (the differences between the mean values of these parameters determined from either Planck or WMAP9 data constitute 2.46$\sigma$ derived from Planck-2013 for $\Omega_{b}h^2$, 2.68$\sigma$ derived from Planck-2013 for $\Omega_{cdm}h^2$ and 2.32$\sigma$ derived from Planck-2013 for $n_s$). On the other hand, the 2$\sigma$ ranges 
from both experiments always overlap. 
\begin{table}[tbp]
\centering
  \caption{The best-fit values ($\mathbf{p}_i$), mean values and 2$\sigma$ confidence limits for parameters of
cosmological models using 4 different observational datasets: WMAP9 {+} HST
{+} BAO {+} SNLS3 ($\mathbf{p}_1$),  WMAP9 {+} HST {+} BAO {+} SN Union2.1 ($\mathbf{p}_2$), Planck {+} HST {+} BAO {+} SNLS3 ($\mathbf{p}_3$),  Planck {+} HST {+} BAO {+} SN Union2.1 ($\mathbf{p}_4$).}
  \medskip\footnotesize{
  \begin{tabular}{|c|c|c|c|c|c|c|c|c|}
    \hline 
    &\multicolumn{2}{c|}{}&\multicolumn{2}{c|}{}&\multicolumn{2}{c|}{}&\multicolumn{2}{c|}{}\\   
Parameters&\multicolumn{2}{c|}{WMAP9+HST}&\multicolumn{2}{c|}{WMAP9+HST}&\multicolumn{2}{c|}{Planck+HST}
&\multicolumn { 2 } { c| } {Planck+HST}\\
&\multicolumn{2}{c|}{+BAO+SNLS3}&\multicolumn{2}{c|}{+BAO+Union2.1}&\multicolumn{2}{c|}{+BAO+SNLS3}
&\multicolumn { 2 } { c| } {+BAO+Union2.1}\\
    &\multicolumn{2}{c|}{}&\multicolumn{2}{c|}{}&\multicolumn{2}{c|}{}&\multicolumn{2}{c|}{}\\
    \cline{2-9}
    &&&&&&&&\\
     &$\mathbf{p}_1$&2$\sigma$ c.l.&$\mathbf{p}_2$&2$\sigma$ c.l.&$\mathbf{p}_3$&2$\sigma$ c.l.&$\mathbf{p}_4$&2$\sigma$
c.l.\\
    &&&&&&&&\\
    \hline
    &&&&&&&&\\
$\Omega_{de}$&0.727&0.722$_{-0.023}^{+0.022}$&0.720&0.718$_{-0.025}^{+0.023}$&0.718&0.719$_{-0.023}^{+0.021}$&
0.721&0.717$_{-0.024}^{+0.023}$\\&&&&&&&&\\
$w_0$&-1.123&-1.120$_{-0.156}^{+0.160}$&-1.114&-1.092$_{-0.190}^{+0.181}$&-1.146&-1.169$_{-0.136}^{+0.139}$
&-1.247&-1.158$_{-0.156}^{+0.165}$\\&&&&&&&&\\
$c_a^2$&-1.171&-1.337$_{-0.288}^{+0.322}$&-1.341&-1.282$_{-0.342}^{+0.731}$&
-1.152&-1.372$_{-0.242}^{+0.235}$&-1.599&-1.374$_{-0.238}^{+0.246}$\\&&&&&&&&\\
10$\Omega_{b}h^2$&0.225&0.225$_{-0.009}^{+0.009}$&0.226&0.225$_{-0.009}^{+0.009}$&0.220&0.221$_{-0.005}^{+0.005}$&
0.220&0.221$_{-0.005}^{+0.005}$\\&&&&&&&&\\
$\Omega_{cdm}h^2$&0.118&0.117$_{-0.006}^{+0.006}$&0.118&0.117$_{-0.006}^{+0.006}$&0.121&0.120$_{-0.004}^{+0.004}$&
0.121&0.120$_{-0.004}^{+0.004}$\\&&&&&&&&\\
$h$&0.718&0.711$_{-0.029}^{+0.028}$&0.709&0.704$_{-0.031}^{+0.032}$&0.713&0.714$_{-0.027}^{+0.027}$&0.718&0.710$_{-0.030
}^{+0.030}$\\&&&&&&&&\\
$n_s$&0.968&0.968$_{-0.021}^{+0.022}$&0.970&0.969$_{-0.022}^{+0.023}$&0.958&0.960$_{-0.012}^{+0.012}$&0.960&0.960$_{
-0.012}^{+0.012}$\\&&&&&&&&\\
$\log(10^{10}A_s)$&3.103&3.096$_{-0.056}^{+0.059}$&3.095&3.097$_{-0.055}^{+0.059}$&3.098&3.089$_{-0.047}^{+0.050}$
&3.090&3.088$_{-0.047}^{+0.050}$\\&&&&&&&&\\
$\tau_{rei}$&0.087&0.086$_{-0.026}^{+0.027}$&0.082&0.087$_{-0.026}^{+0.026}$&0.093&0.089$_{-0.024}^{+0.026}$&
0.089&0.089$_{-0.024}^{+0.026}$\\&&&&&&&&\\
    \hline
  \end{tabular}}
  \label{tab_wpbs}
\end{table}

We now consider the best-fit values, mean values and the 2$\sigma$ marginalized limits for cosmological parameters obtained from CMB data (WMAP9 or Planck) together with HST{+}BAO data. The results are presented in the table \ref{tab_wp} (columns 6-9) and figure \ref{postlike_wp}
(bottom panels). Both WMAP9 and Planck data together with HST{+}BAO prefer a phantom scalar field model of dark energy. Now the dark energy parameters are determined more reliably: the likelihood and posterior functions are similar. The maxima of the marginalized likelihoods and posteriors are close and all ($L(\mathbf{x};w_0)$, $P(w_0;\mathbf{x})$, $L(\mathbf{x};c_a^2)$, $P(c_a^2;\mathbf{x})$) are in the phantom range. Figure \ref{postlike_wp} illustrates also that according to Planck{+}HST{+}BAO data the $\Lambda$CDM model is well outside the 1$\sigma$ contour of $w_0$ and close to the 2$\sigma$ border. We can therefore state that Planck{+}HST{+}BAO dataset disfavors the cosmological constant as dark energy at nearly 2$\sigma$ confidence level. The dataset WMAP9{+}HST{+}BAO also prefers phantom dark energy, but $\Lambda$CDM and quintessence dark energy ($w_0\ge-1$) are within the 1$\sigma$ $\Omega_{de}-w_0$ contour. The 2$\sigma$ confidence ranges of the dynamical dark energy 
parameters $\Omega_{de}$, $w_0$ and $c_a^2$ are now narrower (rows 7 and 9 in table \ref{tab_wp}) and for the WMAP9{+}HST{+}BAO data their full 2$\sigma$ widths are 0.060, 0.506 and 0.979, while for the Planck{+}HST{+}BAO data they are 0.057, 0.387 and 0.486 correspondingly. The 2$\sigma$ confidence ranges for values of the other cosmological parameters are somewhat narrower too; especially, $h$ and $\tau_{rei}$ are now reasonably well determined. The  tension mentioned above for baryon and dark matter density parameters  is significantly less severe: the best-fit values for WMAP9{+}HST{+}BAO dataset are inside  the 2$\sigma$ confidence ranges for Planck{+}HST{+}BAO dataset, the differences between the mean values of these parameters determined with using either Planck or WMAP9 data 
constitute 1.82$\sigma$ derived from Planck{+}HST{+}BAO for $\Omega_{b}h^2$ and 1.58$\sigma$  derived from Planck{+}HST{+}BAO for $\Omega_{cdm}h^2$.

\begin{figure}[tbp]
\includegraphics[width=0.5\textwidth]{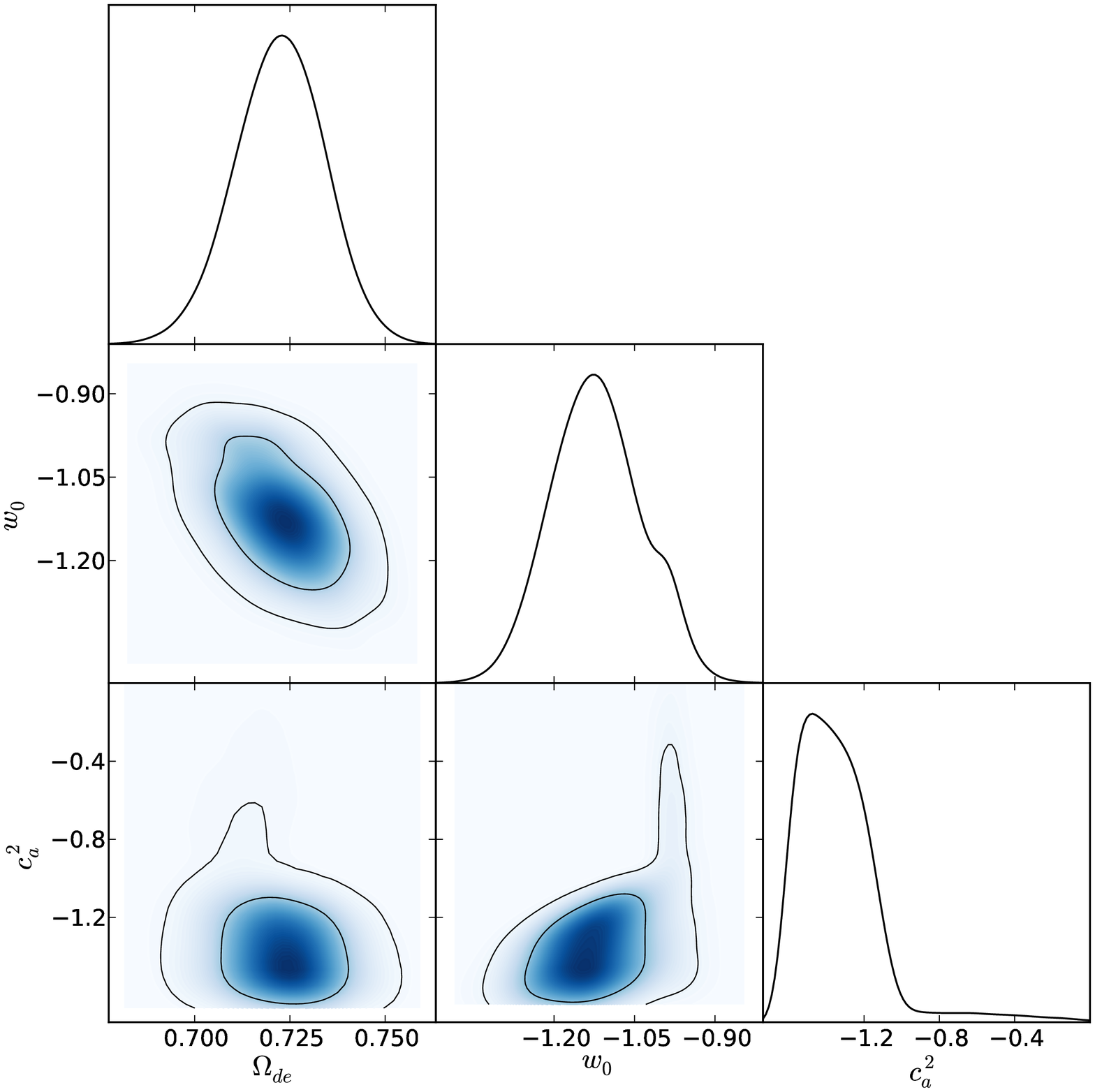}
\includegraphics[width=0.5\textwidth]{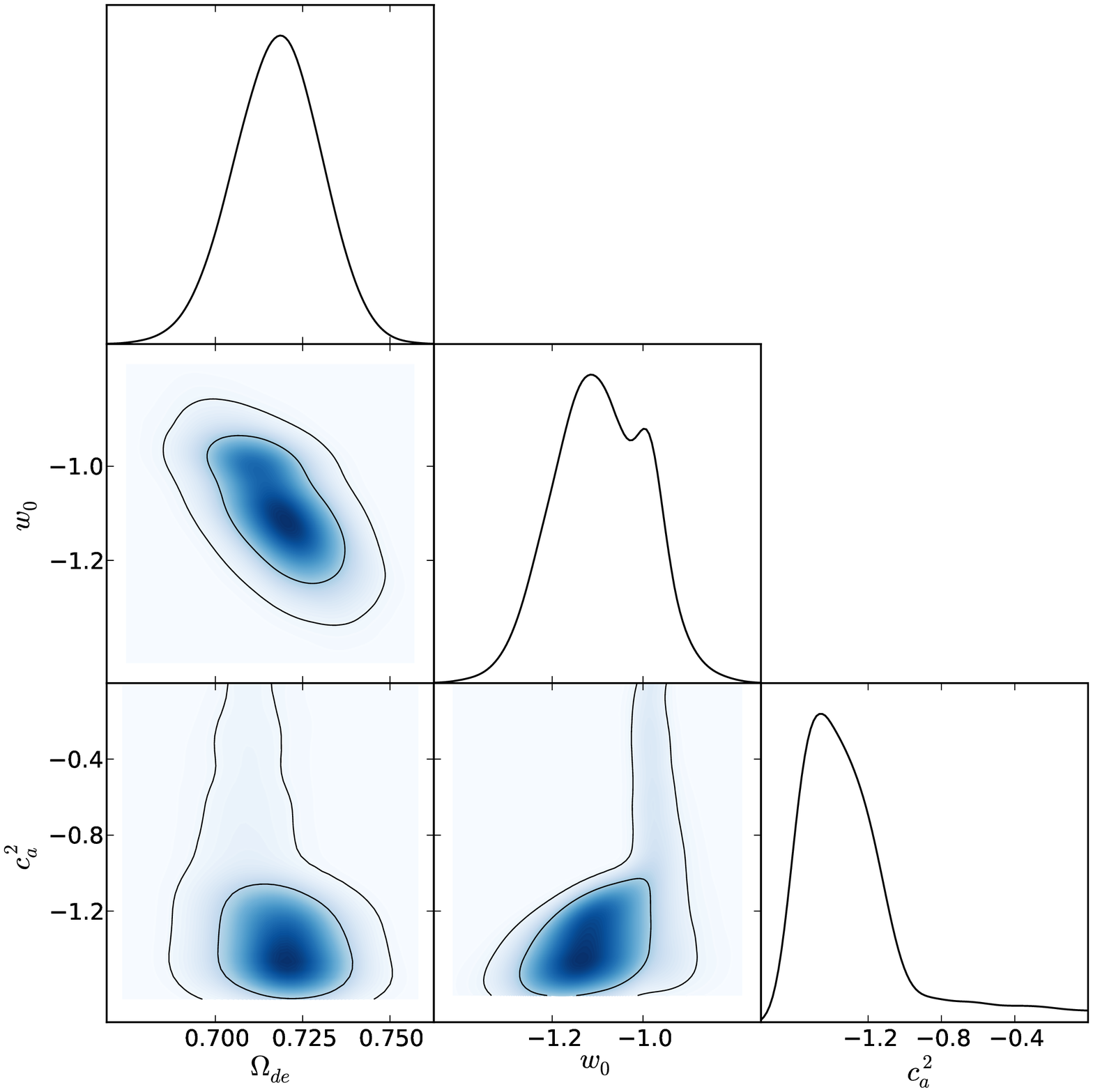}
\includegraphics[width=0.5\textwidth]{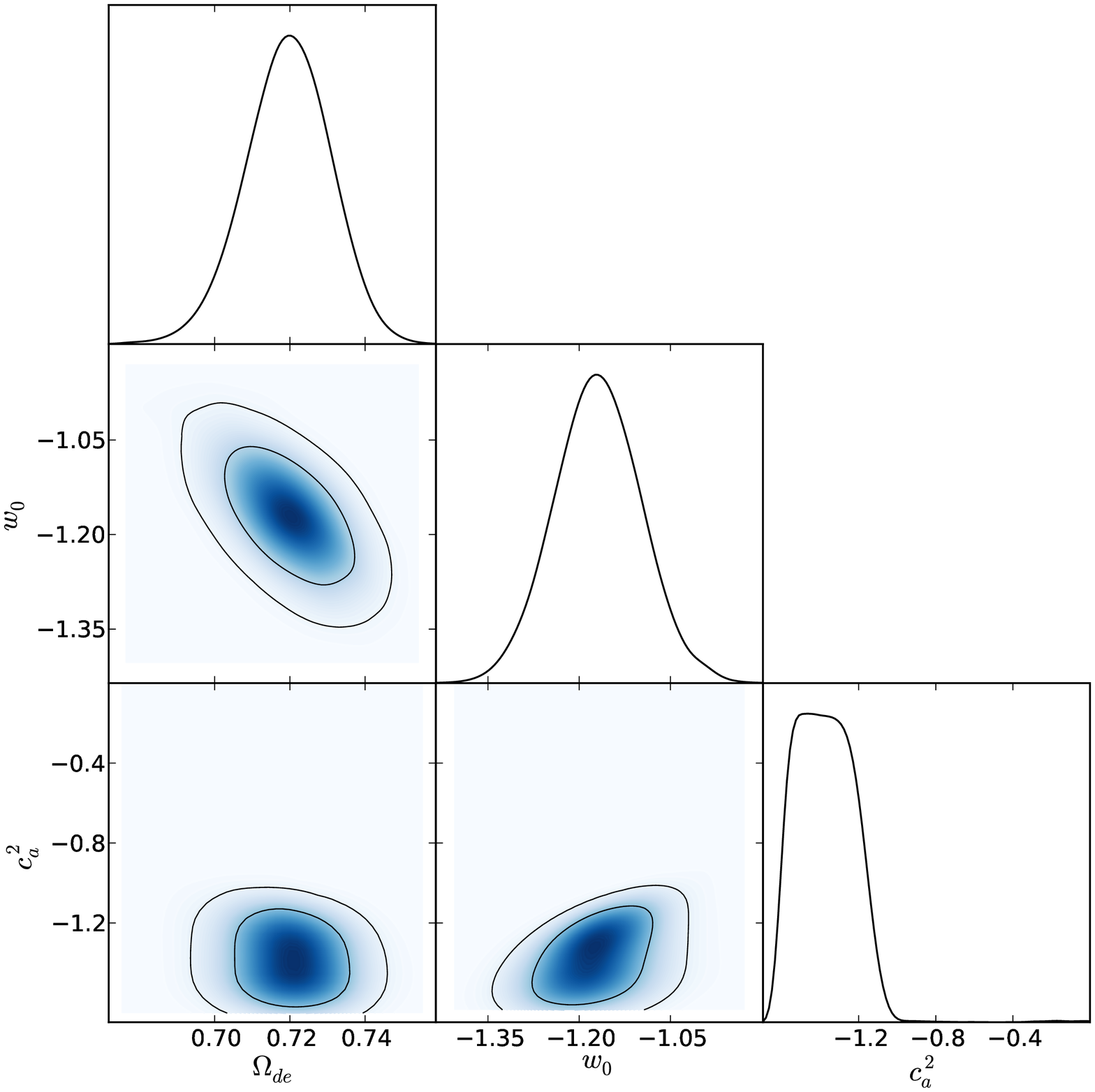}
\includegraphics[width=0.5\textwidth]{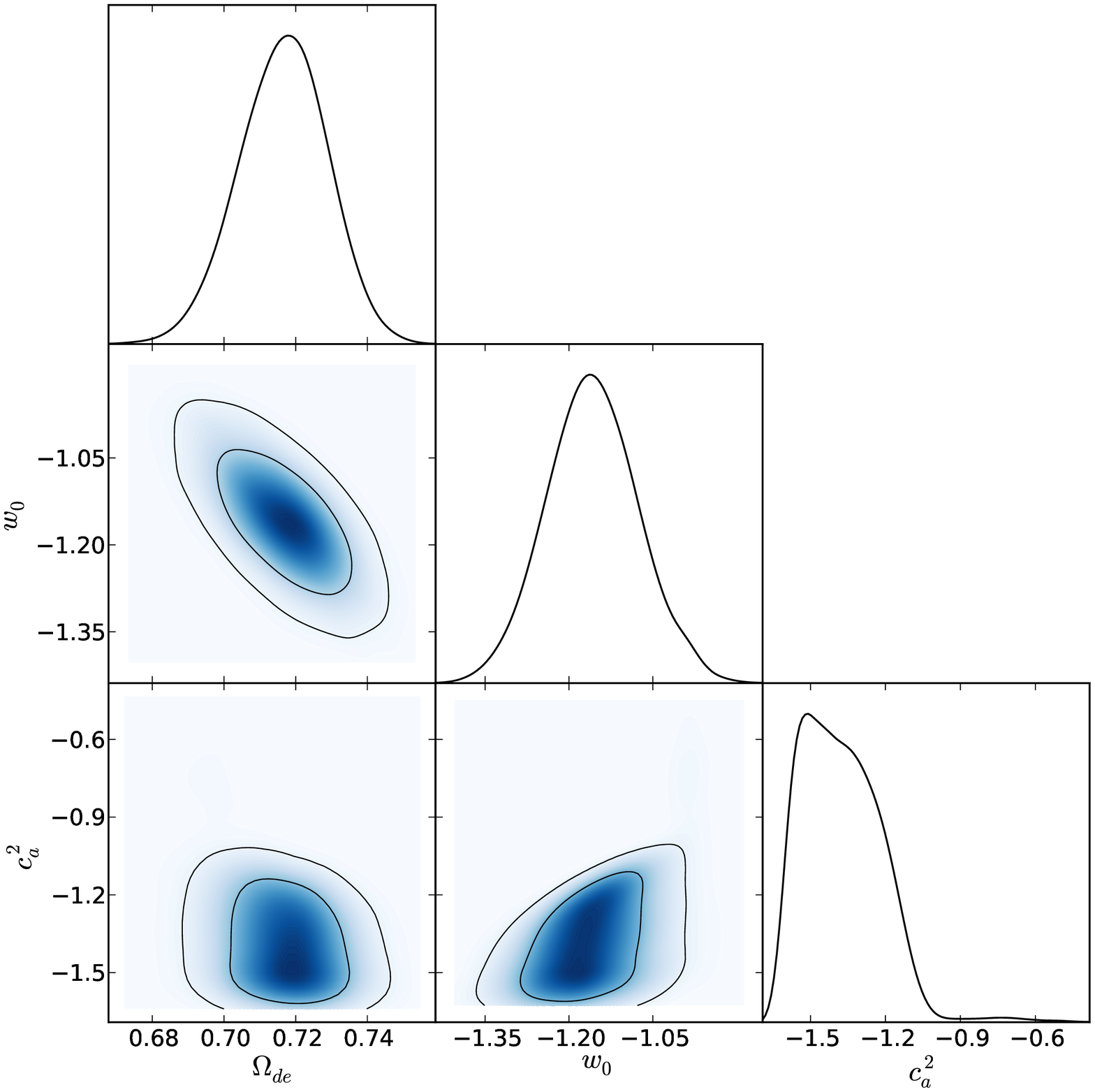}
\caption{One-dimensional marginalized posteriors (solid lines) for $\Omega_{de}$,
$w_0$ and $c_a^2$; color panels show two-dimensional marginalized posterior distributions in the planes $\Omega_{de}-w_0$, $\Omega_{de}-c_a^2$ and $w_0-c_a^2$, where solid lines show the $1\sigma$ and $2\sigma$ confidence contours. Top plots are for WMAP9+HST+BAO data with SNLS3 (left) and Union2.1 (right) SNe Ia compilations; bottom plots are for Planck+HST+BAO with SNLS3 and Union2.1 SNe Ia compilations correspondingly.}
\label{postlike_wpbs}
\end{figure}

The SNe Ia magnitude-redshift relation is the key observational evidence for the existence of dark energy at very high confidence level.  However, some tension exists between distance moduli obtained using different lightcurve fitters applied to the same SNe Ia (for example, SALT2 \cite{Guy2007} and MLCS2k2 \cite{Jha2007}). This has already been highlighted and analyzed in \cite{Kessler2009,Bengochea2011}, but up to now we have no decisive arguments in favor of one of the proposed lightcurve fitters. In Refs.\cite{Bengochea2011,Novosyadlyj2012,Sergijenko2013} it is shown that the dataset with SNe Ia distance moduli determined with the MLCS2k2 fitter prefer quintessence dark energy, while those with SALT2 applied to the same supernovae prefer phantom dark energy. To avoid this ambiguity we use the high-quality joint sample of 472 SNe Ia compiled by \cite{snls3} and denoted here as SNLS3. For these supernovae the updated versions of two independent light curve fitters, SiFTO~\cite{Conley2008} and SALT2~\cite{
Guy2007}, 
have been used for the distance estimations. We denote the two datasets including this compilation by WMAP9{+}HST{+}BAO{+}SNLS3 and Planck{+}HST{+}BAO{+}SNLS3. To evaluate the reliability of the parameters and their confidence intervals based on these datasets we use also other homogeneous sample of distance module - redshift measurements for 580 SNe Ia from the Union2.1 compilation. We denote the datasets with these supernovae  by WMAP9{+}HST{+}BAO{+}Union2.1 and Planck{+}HST{+}BAO{+}Union2.1. The results of the MCMC  determination of  cosmological parameters from these four datasets are presented in  table~\ref{tab_wpbs} and figure~\ref{postlike_wpbs}. We denote the sets of best-fit parameters $\Omega_{de}$, $w_0$, $c_a^2$, $\Omega_bh^2$, $\Omega_{cdm}h^2$, $h$, $n_s$, $A_s$ and $\tau_{rei}$  by $\mathbf{p}_1$, $\mathbf{p}_2$, $\mathbf{p}_3$ and 
$\mathbf{p}_4$ for the WMAP9{+}HST{+}BAO{+}SNLS3,  WMAP9{+}HST{+}BAO{+}SN Union2.1, Planck{+}HST{+}BAO{+}SNLS3 and  Planck{+}HST{+}BAO{+}\\SN Union2.1 datasets respectively.

Let us first compare the cosmological parameters ($\Omega_bh^2$, $\Omega_{cdm}h^2$, $n_s$, $A_s$, $h$, $\tau_{rei}$) from the datasets with SNLS3 and Union2.1 SNe Ia (2, 3 vs 4, 5 and 6, 7 vs 8, 9 columns of table \ref{tab_wpbs}). One finds that for the same non-SN Ia data (CMB{+}HST{+}BAO), the best-fit and mean values of these parameters are practically identical. The confidence limits are different only for the Hubble parameter $h$ and the optical depth to reionization $\tau_{rei}$: they are narrower in determination with SNLS3 moduli distances than with Union2.1. No tension between SNLS3 and Union2.1 data in the determination of the best parameters is found.

Next we compare the results for the  parameters $\Omega_bh^2$, $\Omega_{cdm}h^2$, $n_s$, $A_s$, $\tau_{rei}$ from the WMAP9{+}HST{+}BAO{+}SNe Ia and Planck{+}HST{+}BAO{+}SNe Ia datasets (SNe Ia here denotes either SNLS3 or Union2.1). Clearly, the SNe Ia data does not influence results of these parameter in a significant way: this follows from the comparison of  columns 6 and 7 of table \ref{tab_wp} with  columns 2-5 s of table \ref{tab_wpbs} as well as  columns 
 8 and 9 columns of table \ref{tab_wp} with  columns 6-9 of table \ref{tab_wpbs}. 
We also note that inclusion of SNe Ia data reduces slightly the Planck-WMAP9 tension mentioned above: best-fit values of baryon and dark matter density parameters determined from WMAP9{+}HST{+}BAO{+}SNe Ia are still outside the 1$\sigma$ confidence limits but well inside the 2$\sigma$ range from the Planck{+}HST{+}BAO{+}SNe Ia datasets. The differences between the mean values of $\Omega_{b}h^2$ constitute 1.72$\sigma$ derived from Planck+...+SNLS3 for datasets WMAP9+...+SNLS3 and Planck+...+SNLS3, 1.86$\sigma$  derived from Planck+...+Union2.1 for datasets WMAP9+...+Union2.1 and Planck+...+Union2.1. The differences between the mean values of $\Omega_{cdm}h^2$ are 1.55$\sigma$  derived from Planck+...+SNLS3 for datasets WMAP9+...+SNLS3 and Planck+...+SNLS3, 1.59$\sigma$ derived from Planck+...+Union2.1 for datasets WMAP9+...+\\Union2.1 and Planck+...+Union2.1. The Hubble parameter is determined reliably by the datasets including supernovae: the 2$\sigma$ limits for $h$ 
are now 0.057, 0.063, 0.054 and 0.06 when determined the WMAP9+...+SNLS3, WMAP9+...+Union2.1, Planck+...+SNLS3, Planck+...\\+Union2.1 dataset correspondingly. The dataset Planck{+}HST{+}BAO{+}SNLS3 is the most self-consistent and the most accurate. 

Let us now return to the determination of the dynamical dark energy parameters, see figure \ref{postlike_wpbs}. First, we note that all datasets with SNe Ia moduli distance--redshift relations prefer phantom dark energy. For the WMAP9+...+SNLS3 and WMAP9+...+Union2.1 datas the phantom divide line is within the 1$\sigma$ confidence limits of $w_0$. In the case of Planck+...+Union2.1 dataset it is within the 2$\sigma$ and for the  Planck{+}HST{+}BAO{+}SNLS3 dataset it is outside the  2$\sigma$ confidence limits for $w_0$. Adding the SNe Ia data improves the determination of all the dark energy parameters. This follows from a comparison of the results presented in tables \ref{tab_wp}-\ref{tab_wpbs} and figures \ref{postlike_wp}-\ref{postlike_wpbs}. The 2$\sigma$ confidence ranges of $\Omega_{de}$, $w_0$ and $c_a^2$ are narrowest for the dataset Planck{+}HST{+}BAO{+}SNLS3.
In the models 
with $\mathbf{p}_1$, $\mathbf{p}_2$, $\mathbf{p}_3$ and $\mathbf{p}_4$ parameters the Big Rip singularity (\ref{t_br}) is reached in 73.4, 55.0, 72.4 and 27.8 Gyrs respectively. 

\begin{figure}[tbp]
\centerline{\includegraphics[width=0.5\textwidth]{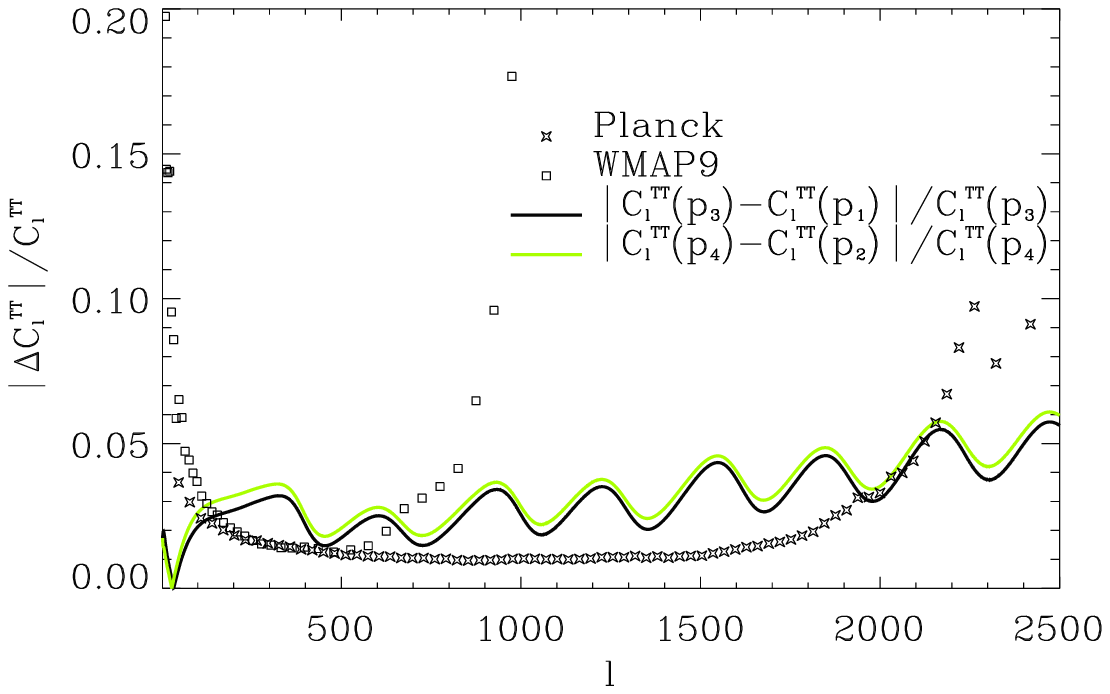}
\includegraphics[width=0.5\textwidth]{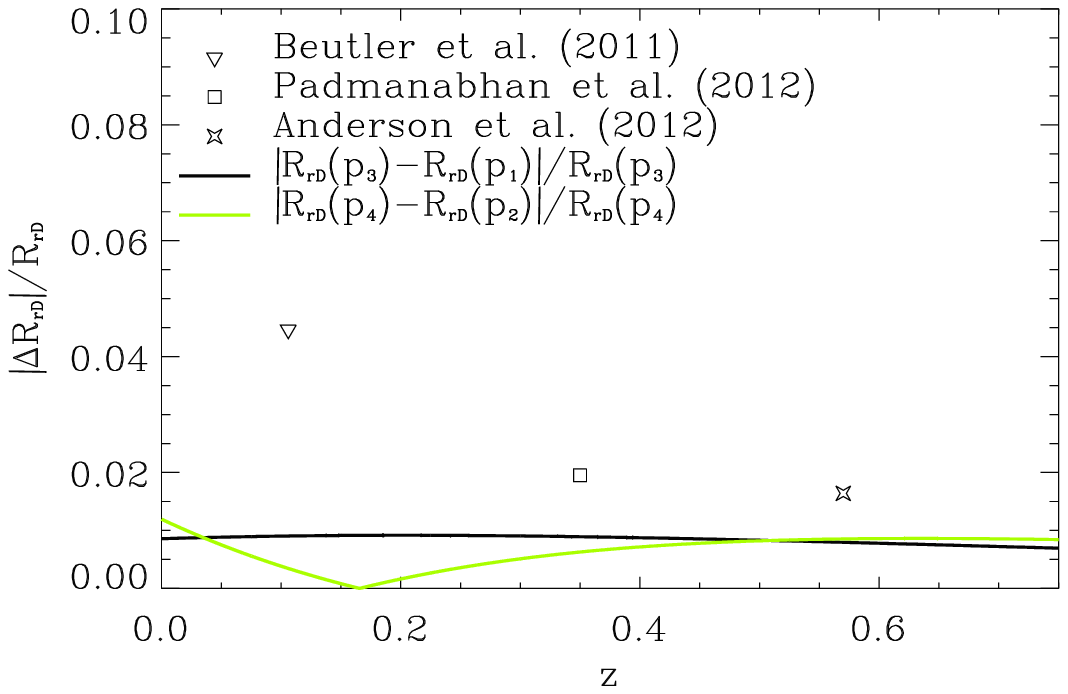}}
\centerline{\includegraphics[width=0.5\textwidth]{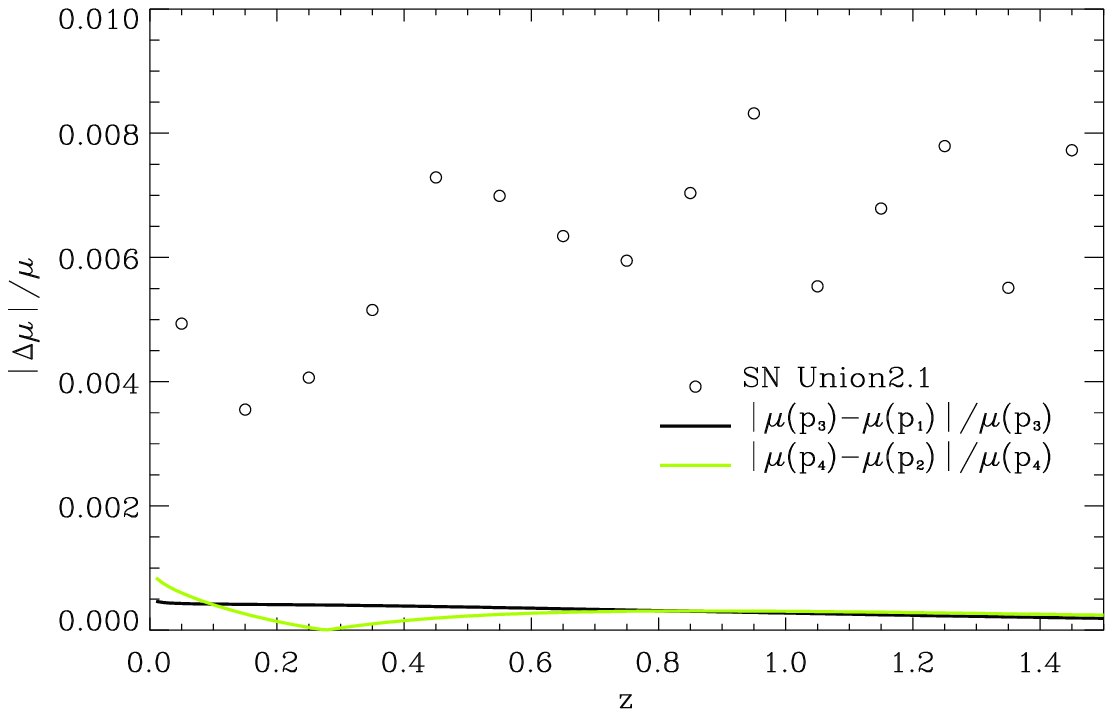}}
\caption{Relative differences of the theoretical predictions from the model with parameters $\mathbf{p}_i$ from table \ref{tab_wpbs} versus the relative uncertainties of observational data for the CMB temperature power spectrum (left top panel), BAO's (right top panel), SNe Ia distance moduli (bottom panel).}
\label{comparison}
\end{figure}
In figure \ref{comparison} we compare the relative differences of the theoretical predictions of models with the parameters $\mathbf{p}_i$ from table \ref{tab_wpbs} with relative uncertainties of observational data on the CMB power spectrum of temperature fluctuations \cite{wmap9a,Planck2013b}, on BAO's \cite{Anderson2012,Padmanabhan2012,6dF}, on SNe Ia distance moduli \cite{union} (binned in 15 bins in $z$ with width of 0.1). We see that the differences in $R_{rD}\equiv r_s(z_{drag})/D_V(z)$ and $\mu$ are smaller than the uncertainties of corresponding data. For $C_{\ell}^{TT}$ the relative differences between models with parameters obtained from Planck and WMAP9 exceed the relative uncertainties of Planck data for $\ell\sim 100-2000$ and the ones of WMAP9 for $\ell\sim 100-650$. This is an additional illustration of the tension between Planck and WMAP9 data.

The results presented in tables \ref{tab_wp} and \ref{tab_wpbs} are in agreement with other determinations, in particular with \cite{wmap9b,Planck2013c,Xia2013,Rest2013,Cheng2013,Shafer2013}. Small differences between the best-fit values of some parameters are due to i) the statistical nature of the MCMC technique, ii) the difference of the dark energy models and iii) differences in the sets of observational data and priors. Our best determination ($\mathbf{p}_3$) gives 
$\Omega_m=0.281\pm0.012$ for the matter density parameter  at 1$\sigma$  and  $w_{de}=-1.169\pm0.069$ for the dark energy EoS parameter today. This means that the dark energy EoS parameter differs from the cosmological constant value of $-1$ by more than 2$\sigma$,  this is in agreement with a similar study in \cite{Rest2013}, which uses the combination of  the 1.5 year Pan-STARRS1 supernovae Ia measurements with Planck{+}HST{+}BAO and 'excludes' the $\Lambda$CDM model of dark energy in a flat Universe at the level of 2.4$\sigma$  (with $\Omega_m=0.277\pm0.012$, $w_{de}=-1.186\pm0.076$).

\section{Conclusion}

We have determined the best-fit values and the confidence limits for parameters of cosmological models with dynamical dark energy using the MCMC technique on the basis of different datasets, which include the results from the final WMAP data release and the Planck-2013 data, the type Ia supernovae samples SNLS3 and Union2.1, the updated BAO measurements together with the HST prior on the Hubble constant. The results, presented in tables \ref{tab_wp}, \ref{tab_wpbs} and figures \ref{postlike_wp}, \ref{postlike_wpbs}, can be summarized as follows:

1) In the class of spatially flat models of the Universe both WMAP and Planck data alone prefer dark energy density dominated models at a very high level of confidence. In the class of thedynamical dark energy models  studied here, WMAP9 data alone prefer quintessence, but $\Lambda$ and phantom models are within 1$\sigma$  of $w_{de}$. The Planck-2013 data alone, in contrary, prefer phantom models of dark energy, but the confidence level of this preference is low, $\Lambda$ and quintessence models are within the 1$\sigma$ confidence limits for $w_{de}$. The confidence limits of the dark energy parameters are narrower for the Planck data than for WMAP9.

2) Adding HST and BAO data to WMAP9 and Planck-2013 data improves the accuracy of the determinations of $\Omega_bh^2$, $\Omega_{cdm}h^2$, $n_s$ and $\log(10^{10}A_s)$
Both WMAP9 and Planck data together with HST{+}BAO prefer a phantom scalar field model of dark energy. For WMAP9{+}HST{+}BAO, the $\Lambda$CDM model and quintessence dark energy ($w_0\ge-1$) are inside the 1$\sigma$ confidence limits of $w_0$, while for Planck{+}HST{+}BAO the $\Lambda$ model is outside the 1$\sigma$ confidence limits of $w_0$ but still inside the  2$\sigma$ range. For the Planck{+}HST{+}BAO dataset the confidence limits for the dynamical dark energy parameters $\Omega_{de}$, $w_0$ and $c_a^2$ are significantly narrower.

3) Adding finally supernova data, the SNe Ia samples SNLS3 or Union2.1, to WMAP9 {+}HST{+}BAO or Planck{+}HST{+}BAO 
increases the precision of the Hubble constant and of the dynamical dark energy parameters. In all combinations of WMAP9{+}HST{+}BAO and Planck{+}HST{+}BAO datasets with SNLS3 and Union2.1, the phantom scalar field model of dark energy is preferred. The most reliable determination of cosmological and dynamical dark energy parameters is obtained from the Planck{+}HST{+}BAO{+}SNLS3 dataset. The best-fit values of the parameters and their 2$\sigma$ confidence limits  are: $\Omega_{de}=0.718\pm0.022$, $w_0=-1.15^{+0.14}_{-0.16}$, $c_a^2=-1.15^{+0.02}_{-0.46}$, $\Omega_bh^2=0.0220\pm0.0005$, $\Omega_{cdm}h^2=0.121\pm0.004$, $h=0.713\pm0.027$,  $n_s=0.958^{+0.014}_{-0.010}$, $A_s=(2.215^{+0.093}_{-0.101})\cdot10^{-9}$, $\tau_{rei}=0.093^{+0.022}_{-0.028}$. The $\Lambda$CDM model is disfavored by this dataset at 2$\sigma$ confidence. The dataset WMAP9{+}HST{+}BAO{+}SNLS3 disfavors the $\Lambda$-term only at 1$\sigma$.

4) The results presented in the tables \ref{tab_wp} and \ref{tab_wpbs} highlight a tension between WMAP9 and Planck-2013 alone: the best-fit and mean values of the baryon and dark matter density parameters as well as  the spectral index determined from datasets including WMAP9 are outside the 2$\sigma$ limits of the corresponding values determined from datasets with Planck-2013. When including the HST+BAO+SNIa data this tension is reduced below 2$\sigma$ in all parameters. Note, however that the tension in the Hubble parameter from WMAP and Planck data present for the $\Lambda$CDM model disappears in our dynamical dark energy model.

The CMB power spectra, the BAO distance ratios and the SN Ia distance moduli computed for the cosmological models with best-fit parameters $\mathbf{p_i}$ (table \ref{tab_wpbs}) match well observational data that have been used in the search procedure.

\acknowledgments

This work was supported by the project of Ministry of Education and Science of Ukraine (state registration number 
0113U003059), research program ``Scientific cosmic research'' of the National Academy of Sciences of Ukraine (state registration number 0113U002301) and the SCOPES project No. IZ73Z0128040 of the Swiss National Science Foundation. Authors also acknowledge the use of CAMB and CosmoMC packages.

\end{document}